\documentstyle[11pt,epsfig]{article}

\newcommand{\spa}{\vspace{.25cm}}   
\newcommand{\beq}{\begin{equation}}
\newcommand{\eeq}{\end{equation}}
\newcommand{\beqs}{\begin{eqnarray}}
\newcommand{\eeqs}{\end{eqnarray}}
\newcommand{\bary}{\begin{array}}
\newcommand{\eary}{\end{array}}
\newcommand{\nid}{\noindent}
\newcommand{\sni}{\spa \nid}
\newcommand{\e}{e}
\newcommand{\x}{\ell}
\newcommand{\FCNC}{FCNCs }

\newcommand{\step}{\hspace{1.4cm}}
\newcommand{\To}{\Longrightarrow}
\newcommand{\figpos}{p}        
\newcommand{\spaContour}{4.7}  
\newcommand{\heightCont}{16.2} 
\newcommand{\widthCont}{16.2}  
\newcommand{\spaAR}{7.5}       
\newcommand{\heightAR}{16.2}   
\newcommand{\widthAR}{16.2}    

\def\gsim{\ \rlap{\raise 3pt \hbox{$>$}}{\lower 3pt \hbox{$\sim$}}\ }
\def\lsim{\ \rlap{\raise 3pt \hbox{$<$}}{\lower 3pt \hbox{$\sim$}}\ }

\def\putMSW#1#2#3#4#5 
{
 \begin{figure}[\figpos]
 \LARGE
 $\Delta$ [eV$^2$]
 \begin{center}
 \mbox{\epsfig{figure=#1.eps,angle=0,width=#3cm,height=#4cm}}
 \end{center}
 \hspace{#5cm}
 $\sin^2 2\theta$
 \spa
 \caption{#2}
 \label{#1}
 \end{figure}
}

\begin{document}

\title{The Solar Neutrino Problem in the Presence of \\
       Flavor Changing Neutrino Interactions}

\author{Sven Bergmann\\ 
        \small \it Department of Particle Physics,
        Weizmann Institute of Science,
        Rehovot 76100, Israel}

\date{(WIS-97/19/Jul-PH)}
\maketitle

\begin{abstract}%
We study the effects of flavor changing neutrino interactions
on the resonant conversion of solar neutrinos. In particular, we
describe how the regions in the $\Delta m^2-\sin^22\theta$ plane
that are consistent with the four solar neutrino experiments are modified
for different strengths of New Physics neutrino interactions. 
\end{abstract}%

\hspace{4cm}%


\section{Introduction}
Neutrino oscillations \cite{MSW, Parkes, KuoPanta, Barger91a} are
considered to be the most likely solution to the longstanding Solar
Neutrino (SN) Problem \cite{Bahcall, Langack94, Langack97}. The
standard solution asserts that neutrinos have non-vanishing masses and
that there is mixing. Many extensions of the Standard Model (SM), such
as Left-Right Symmetric Models (LRSMs) \cite{LRSM} and Supersymmetric
Models without $R$-parity \cite{SUSY}, predict not only neutrino
masses but also New Physics (NP) neutrino interactions that are not
present in the SM.

\spa%
The effects of NP on neutrino oscillations have been studied
previously in the literature \cite{Roulet, Guzzo, Barger91b, Fogli,
  KB97}.  In~\cite{Roulet} the effects of flavor changing neutrino
interactions on the Mikheyev-Smirnov-Wolfenstein (MSW) effect
\cite{MSW} were considered and it was demonstrated that, even in the
absence of neutrino mixing in vacuum, neutrino oscillations in matter
can be enhanced by flavor changing neutral currents (FCNCs).
In~\cite{Guzzo} it was shown that additional non-universal flavor
diagonal neutral currents (FDNCs) may allow for resonantly enhanced
neutrino transitions even if neutrinos have vanishing masses. Both
scenarios as well as the case of massive neutrinos with off-diagonal
and new diagonal currents were investigated thoroughly in
Ref.~\cite{Barger91b}. This analysis also determined the regions in
parameter space that gave consistency with the standard solar model
(SSM) and the then available solar neutrino data from the Homestake
\cite{HS} and Kamiokande \cite{KK} experiments.  The implications on
the allowed regions due to the more recent data from the two gallium
detectors, GALLEX \cite{GALLEX} and SAGE \cite{SAGE}, were discussed
in \cite{Fogli} for massive neutrinos and FCNCs (but no FDNCs).  More
recently also the solution to the SN Problem with both FCNCs {\it and}
FDNCs but with massless neutrinos was re-analyzed \cite{KB97}, this
time including the data from all four SN experiment and the latest
improvements to the SSM \cite{BP95}.

\spa%
In this work we reconsider the combination of neutrino masses and
mixing with new flavor changing neutrino interactions (but without
significant non-universal FDNCs). We update the previous analyses of
this case by taking into account the most recent data from all four SN
experiments and by using the latest improvements to the SSM. We
present the updated combined allowed regions of the four SN
experiments for different strengths of the NP neutrino couplings. In
particular we include the effects due to the variation of the relative
NP strength that arises when the solar neutrinos scatter off quarks on
their way to the solar surface.  We show that for a certain range of
the NP coupling it is possible to solve the SN problem with
vanishingly small vacuum mixing. Moreover we reveal some interesting
analytic details of the MSW resonant conversion of solar neutrinos in
the presence of FCNCs.

\spa%
The paper is structured as follows: In Section~\ref{formalism} we
introduce the formalism of neutrino oscillations with FCNCs. In
Section~\ref{lepFCNC} we explore the effects on the MSW resonant
conversion for purely leptonic NP and use our results in
Section~\ref{lepNPexps} to plot the MSW-contours for the three types
of SN experiments.  From this we obtain the combined allowed regions
for various NP-strengths in this sector.  In Section~\ref{quarkFCNC}
we discuss the new features arising when the neutrinos scatter off
quarks and we present the MSW-contours and the new allowed regions in
Section~\ref{quarkNPexps}.  Finally we conclude in
Section~\ref{conclusions} and discuss briefly how the NP couplings we
require are constrained by SM-forbidden decays.


\section{Formalism} 
\label{formalism}

NP interactions may affect the neutrino propagation through matter.
In particular, the resonant conversion of electron neutrinos produced
in the center of sun is modified in the presence of FCNC neutrino 
scattering off electrons and nucleons.

\spa%
To a good approximation the equation of motion for two neutrino
flavors $\nu_e$ and $\nu_\x$ ($\x=\mu, \tau$) in the presence of
matter induced \FCNC is given by \cite{Roulet}

\beq
i\frac{d}{dt} \pmatrix{\nu_e \cr \nu_\x} = 
{\cal{H}}_N \pmatrix{\nu_e \cr \nu_\x} =
\frac{1}{4E}
\pmatrix{-\Delta \cos2\theta + A & \Delta \sin 2\theta + B \cr
          \Delta \sin2\theta + B & \Delta \cos 2\theta - A \cr} 
\pmatrix{\nu_e \cr \nu_x},
\label{eq-of-motion} 
\eeq

\sni%
where $E$ is the neutrino energy, $\Delta \equiv m^2_2 - m^2_1$ is the
mass-squared difference of the two vacuum mass-eigenstates and
$\theta$ is the vacuum mixing angle. (Note that in the presence of
non-standard neutrino interactions the weak eigenstates are not always
flavor eigenstates \cite{Yuval}.)  $A \equiv 2E \sqrt{2} G_F N_e$ is
the standard induced mass due to $W$-exchange in the reaction $\nu_e
\e \to \nu_e \e$ and the parameter

\beq
B \equiv 4E \sqrt{2} (G_N^e N_e + G_N^u N_u + G_N^d N_d)
\eeq 

\sni%
describes the FCNC contributions from neutrino scattering off
electrons and quarks in the sun. Here $N_f$ denotes the number density of
the fermion type $f$ and $G_N^f$ is the effective four-Fermi coupling
of the reaction $\nu_e f \to \nu_\x f$. It is convenient to rewrite
this as

\beq
B  = 4E \sqrt{2} G_F N_e \left[\epsilon_e 
          + 2 \epsilon_u + \epsilon_d 
          + \left(\epsilon_u + 2\epsilon_d \right) \frac{N_n}{N_e} \right].
\label{Bdef}
\eeq

\sni%
We introduced the parameters 

\beq
\epsilon_f \equiv G_N^f / G_F
\eeq

\sni%
and used the fact that the quark densities can be expressed in terms
of the neutron density $N_n$ and the electron density $N_e$ which
equals to the proton density $N_p$ for neutral matter like in the sun.

\spa%
In the following section we will first analyze the case where only
$\epsilon_e$ is non-vanishing. This case is the simplest and displays
most of the features we will encounter later when discussing the more
complicated case of \FCNC from scattering off quarks.


\section{\FCNC in the Leptonic Sector}
\label{lepFCNC}

Assume that the NP relevant to the neutrino propagation appears only
in the leptonic sector, i.e.  $\epsilon_e \gg
\epsilon_d, \epsilon_u$.  When rewriting the Hamiltonian ${\cal{H}}_N$
in the equation of motion (\ref{eq-of-motion}) for the neutrino
propagation with matter induced \FCNC as

\beq
{\cal{H}}_N =
\frac{\Delta_N}{4E}
\pmatrix{-\cos2\theta_N & \sin 2\theta_N \cr
         \sin2\theta_N & \cos 2\theta_N \cr} 
\eeq

\sni%
we obtain that the effective mixing is given by

\beq
\cos2\theta_N = \frac{(\Delta \cos2\theta - A)}
  {\sqrt{(\Delta \cos2\theta - A)^2 + (\Delta \sin2\theta + B)^2}}
\label{cos-theta_N}
\eeq

\sni%
and the effective mass-squared difference is

\beq
\Delta_N = \frac{\Delta \sin 2\theta + B}{\sin 2\theta_N}.
\eeq

\spa%
For the range of the parameters $\Delta$ and $\theta$ relevant to the
MSW-effect there are typically many oscillations between the neutrino
production and a resonance, and again between the resonance and detection. 
Hence the phase information from before and after resonance is easily 
lost. In this case one may use the averaged probability for a neutrino 
produced in the solar center to be detected as an electron neutrino 
which is given by \cite{Parkes, KuoPanta}

\beq
P_N(\nu_e \to \nu_e) = 
 \frac{1}{2} + (\frac{1}{2} - P_c) \cos 2\theta ~\cos 2\theta_N.
\label{Pee}
\eeq

\sni%
If a neutrino is produced above the resonance ($A_{prod} \ge
A_{res}=\Delta \cos 2\theta$) then level-crossing can occur. This is
accounted for by the crossing probability $P_c$ in (\ref{Pee}) which
is well approximated by~\cite{Petcov, KuoPanta}

\beq
P_c=\Theta(A_{prod}-A_{res}) \times 
    \frac{\exp \left[\pi \gamma_N F(\theta)/ 2 \right]-
          \exp \left[\pi \gamma_N F(\theta)/ 2 \sin^2 2\theta \right]}
         {1 - \exp \left[\pi \gamma_N F(\theta)/ 2 \sin^2 2\theta \right]}.
\label{Pc}
\eeq

\sni%
If we assume that the electron density in the sun $N_e \propto
\exp(-r)$ then the parameter $F$ takes the value $F(\theta)=1-\tan^2
\theta$. The adiabaticity parameter $\gamma_N$ in (\ref{Pc}) is
defined as

\beqs
\gamma_N &\equiv& \left|\frac{\Delta_N/4E}{d\theta_N/dx}\right|_{res} 
 \nonumber \\
 &=& \left|\frac{(\Delta \sin 2\theta + B_{res})}{4E}
           \frac{2 \Delta (\sin 2\theta + 2\epsilon_e \cos 2\theta)}
                {A_{res} ~(dN_e/dx)/N_e|_{res}} \right| \nonumber \\
 &=&  \gamma |1+2\epsilon_e \cot 2\theta|^2,
\label{gamma_N}
\eeqs

\sni%
where the standard adiabaticity parameter is $\gamma=\frac{\Delta
  \sin^2 2\theta} {2 E \cos2\theta ~(dN_e/dx)/N_e|_{res}}$.  
We used

\beqs
\frac{d{\theta}_N}{dx} &=& 
     {\left(\frac{d{\tan 2\theta}_N}{d{\theta}_N}\right)}^{-1}
      \left(\frac{d{\tan 2\theta}_N}{dA}\right)
      \left(\frac{dA}{dx}\right) \nonumber \\
&=& \left(\frac{\cos^2 2\theta_N}{2}\right)
    \left(\Delta \frac{\sin 2\theta +
        2\epsilon_e \cos 2\theta}{(\Delta \cos 2\theta - A)^2}\right)
    \left(\frac{A}{N_e} \frac{dN_e}{dx} \right) \nonumber \\
&=& \frac{1}{2} \frac{\Delta (\sin 2\theta + 2\epsilon_e \cos 2\theta)}
        {(\Delta \cos 2\theta - A)^2 + (\Delta \sin 2\theta + B)^2}
    \left(\frac{A}{N_e} \frac{dN_e}{dx} \right)
\eeqs    

\sni%
and $\Delta_N|_{res} = \Delta \sin 2\theta + B_{res}$.  Note
that for purely leptonic \FCNC the induced mass at the
resonance $A_{res} = \Delta \cos 2\theta$ is linear to $B_{res} =
2\epsilon_e \Delta \cos 2\theta$.

\spa%
From the expression (\ref{gamma_N}) for $\gamma_N$ one can see that
for $\epsilon_e \gsim \tan 2\theta$ there is a considerable modification
to the standard adiabaticity parameter $\gamma$.  Since the standard
non-adiabatic threshold energy 
$E_{NA}=\frac{\pi \Delta \sin^2 2\theta}
             {4 \cos2\theta ~(dN_e/dr)/N_e)|_{res}}$ 
is proportional to the
adiabaticity parameter it has to be corrected by the same factor
\mbox{$|1+2\epsilon_e \cot2\theta|^2$}:

\beq
E_{NA}(\epsilon_e)= 
\frac{\pi \Delta \sin^2 2\theta \times |1+2\epsilon_e \cot2\theta|^2}
     {4 \cos2\theta ~(dN_e/dr)/N_e)|_{res}}.
\label{E_NA_eps}
\eeq

\spa%
We have plotted the survival probability $P_N(\nu_e \to \nu_e)$
of Eq.~(\ref{Pee}) in the $\Delta - \sin^2 2\theta$ plane for a fixed
energy $E$ and different values of $\epsilon_e$. (See
Fig.~\ref{NP-monop} for positive and Fig.~\ref{NP-monom} for negative
$\epsilon_e$, the shading indicates the value of $P_N$: White
corresponds to $0.9 \le P_N \le 1.0$ and the darkest area
corresponds to $0.0 \le P_N \le 0.1$.) For $|\epsilon_e|$ as
small as $0.001$ the effects of NP are minor (compared to any standard
MSW-plot).  However already for $|\epsilon_e| = 0.01$ the triangular
shape is distorted and the originally diagonal band appears bent and
-- most striking -- has a ``gap" for negative $\epsilon_e$.  For even
larger $|\epsilon_e|$ these features remain, only the gap moves
towards larger $\sin^2 2\theta$. Our goal is to understand why these
changes arise.

\spa%
Note that for a given $E_{NA}$ and small vacuum mixing angle we have
\mbox{$\Delta \propto 1/|\sin 2\theta + 2\epsilon_e \cos 2\theta|^2$}.
For $\epsilon_e = 0$ this yields the diagonal contours in a
double-logarithmic plot in the $\Delta - \sin^2 2\theta$ plane
known as the non-adiabatic band.  However for non-vanishing 
$\epsilon_e$ these contours are not diagonal, since $\Delta$ is not
proportional to $\sin^2 2\theta$ anymore. For $\epsilon_e > 0$ the 
contours are below the original diagonals and approach a constant 
for $\sin^2 2\theta \to 0$. For $\epsilon_e < 0$ the behavior of the 
non-adiabatic band is more complicated. For large mixing the new 
contours are above the standard diagonal and diverge at

\beq
\tan 2\theta_{div} = -2\epsilon_e,
\label{tan_div}
\eeq

\sni%
where the correction-factor $|1+2\epsilon_e \cot2\theta|^2$ vanishes and
hence also $E_{NA}(\epsilon_e) = \gamma_N = 0$. This implies that -- 
independently of $\Delta$ -- almost all the electron neutrinos 
which were produced above the resonance mainly as the heavier
mass eigenstate  will ``cross over" at the resonance to the lighter mass 
eigenstate and therefore leave the sun as electron neutrinos. Hence the 
survival probability is very large, which explains why the contours 
split at $\sin^2 2\theta_{div}$. Due to the absolute value in 
Eq.~(\ref{E_NA_eps}), the contours are symmetric in the region
around $\sin^2 2\theta_{div}$. For very small vacuum mixing $\sin^2
2\theta \ll \sin^2 2\theta_{div}$ the contours approach the same 
constant values as in the case of positive $\epsilon_e$. 

\begin{figure}[\figpos]
\LARGE
$P_N(E)$
\begin{center}
\mbox{\epsfig{figure=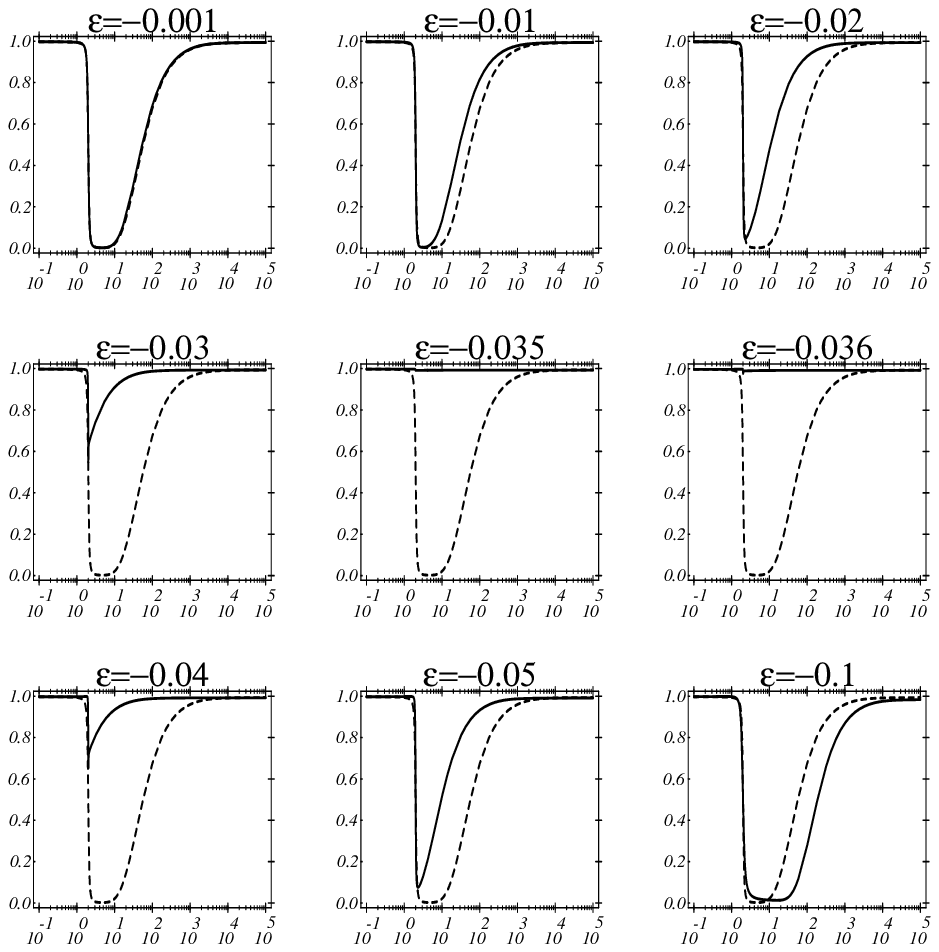,angle=0,width=16.2cm,height=16.2cm}}
\newline
\mbox{~~~~}$E$~[MeV]
\spa
\end{center}
\caption{The survival probability $P_N(E)$ for different 
  $\varepsilon=\epsilon_e$ ($\sin^2 2\theta = 0.005$, $\Delta = 3
  \times 10^{-5} \rm eV^2$). The solid curves correspond to
  $\varepsilon$ as indicated above the plot and the dashed curves
  correspond to $\varepsilon = 0$.}
\label{NP-probs}
\end{figure}

\spa%
To understand how the contours split it is instructive to display
the survival probability (\ref{Pee}) as a function of the energy $E$
for fixed $\sin^2 2 \theta$ and $\Delta$ as shown in
Fig.~\ref{NP-probs}. The solid curves show $P_N(E)$ for some negative
$\epsilon_e$, while the dashed curves indicate $P_N(E)$ for
$\epsilon_e = 0$. Note that when $\epsilon_e$ approaches the value
where $\sin^2 2\theta \simeq \sin^2 2\theta_{div}$, the ``valley"
disappears -- due to the decrease of the non-adiabatic threshold
$E_{NA}(\epsilon_e)$ -- and hence almost all electron neutrinos
survive. Also note that the left side of the valley corresponding to
the adiabatic threshold is almost unaffected by NP interactions. Thus
the adiabatic (horizontal) solution is not shifted.  Moreover if
the vacuum mixing is large ($\sin^2 2\theta \approx \tan^2 2\theta \gg 
\epsilon_e$) then the NP correction factor becomes negligible 
($|1+2\epsilon_e \cot2\theta|^2 \simeq 1$) with the result that the 
large-angle solution is inert to NP.


\section{Solar Neutrino Experiments and $\nu_e-\e$ \FCNC}
\label{lepNPexps}
Once we have found the new survival probability $P_N(E)$ [see
(\ref{Pee})], it is straightforward to predict the effects of NP on
the MSW-plots for the three kinds of SN experiments in the presence of
NP.  In order to obtain the suppression rate $\rho$ at any point in the
$\Delta - \sin^2 2\theta$ plane the survival probability $P_N(E)$ has
to be convoluted with the neutrino energy spectrum times the
sensitivity of each experiment. (For all plots we use the SSM predictions 
of Ref.~\cite{BP95} (including Helium and heavy metal fusion) 
and the experimental results as summarized in Table~I of
Ref.~\cite{Langack97}. We neglect day-night effects \cite{daynight}.) 
Fig.~\ref{NP_all_e} shows the contours for
$\epsilon_e = \pm 0.05$ for Kamiokande, Homestake and the Gallium
experiments, respectively. The plots exhibit basically the same
behavior as those we showed for a discrete energy (Fig.~\ref{NP-monop}
and Fig.~\ref{NP-monom}). The only difference is the distortion
produced by the energy spectrum, as known from the standard 
MSW-effect.

\spa%
Finally we have plotted the allowed regions determined by the ratio $\rho$
between the expected and measured neutrino fluxes for the three types
of SN experiments. We present the individual 95\% C.L. contours (dotted for 
the combined gallium experiments, dashed for the Homestake and solid for the 
Kamiokande experiment) together with the {\it combined} allowed regions 
(shaded) for positive and negative $\epsilon_e$ in Fig.~\ref{AR_NP_pe} and 
Fig.~\ref{AR_NP_me}, respectively. For very small $\epsilon_e = \pm 0.001$ 
and $\epsilon_e = \pm 0.01$ the small-angle solution appears a little 
shifted, but it does {\it not} disappear.  Note that the NP neutrino 
interactions do not change the allowed value for the mass difference, which 
is fixed at $\Delta \approx 10^{-5} \rm eV^2$ due to the adiabatic solution 
of the Gallium experiments.  For larger values $\epsilon_e = \pm 0.05$
and $\epsilon_e = \pm 0.1$ the effects are more dramatic.  One can see
that for $\epsilon_e = \pm 0.05$ the combined allowed regions include
{\it all} vacuum mixing angles for which $\sin^2 2\theta \lsim 3
\times 10^{-4}$. This means in particular that in the presence of NP
interactions there may be a solution to the SN problem with
vanishingly small vacuum mixing.  Moreover for $\epsilon_e = -0.05$
there appears a second solution at $\sin^2 2\theta \simeq 3 \times
10^{-2}$, which is due to the gap for negative $\epsilon_e$.  Finally
for $\epsilon_e = +0.1$ there is {\it no} overlap of the small-angle
allowed regions, while for $\epsilon_e = -0.1$ there are two solutions
at $\sin^2 2\theta \simeq 10^{-2}$ and $\sin^2 2\theta \simeq 7 \times
10^{-2}$ on both sides of the gap. Moreover -- as we expected -- the
large-angle solution ($\sin^2 2\theta \simeq 0.6-1.0$ and $\Delta
\simeq 3 \times 10^{-6}-10^{-4}~{\rm eV}^2$) is inert to NP for all
$\epsilon_e$.


\section{\FCNC in the Quark Sector}
\label{quarkFCNC}

So far we have restricted our discussion to the case of purely
leptonic interactions. However electron neutrinos which have been
produced in the core of the sun may as well scatter off protons and
neutrons when propagating to the solar surface. Without NP
interactions this only gives rise to additional weak neutral currents
via $Z$-exchange which do not alter the resonant conversion. But in
the presence of NP there can be FDNCs and FCNCs which may affect the
neutrino propagation. We only consider flavor changing neutrino 
interactions which correspond to non-vanishing $\epsilon_u$ and/or 
$\epsilon_d$ in~(\ref{Bdef}). In this case a new parameter, the ratio $R 
\equiv N_n/N_p$ plays a role. If $R$ were constant, we would only need to 
replace $\epsilon_e$ of Section~\ref{lepFCNC} by

\beq
\epsilon(R) = \epsilon_e + 2 \epsilon_u + \epsilon_d +
(\epsilon_u + 2\epsilon_d)~R
\label{eps_R}
\eeq

\sni%
and could then use all the results we obtained so far.  However $R$ is
in fact {\it not}\/ a constant, but changes from a value of about
$0.48$ at the center of the sun to $0.16$ at its surface according to
the SSM predictions~\cite{Bahcall}. Taking into
account that $^4$He is four times heavier than $^1$H and neglecting
the contributions from heavier elements, whose abundances are much
smaller, we obtain that

\beq
R \simeq \frac{X(^4{\rm He})}{2X(^1{\rm H})+X(^4{\rm He})}.
\eeq

\begin{figure}[\figpos]
\begin{center}
\mbox{\epsfig{figure=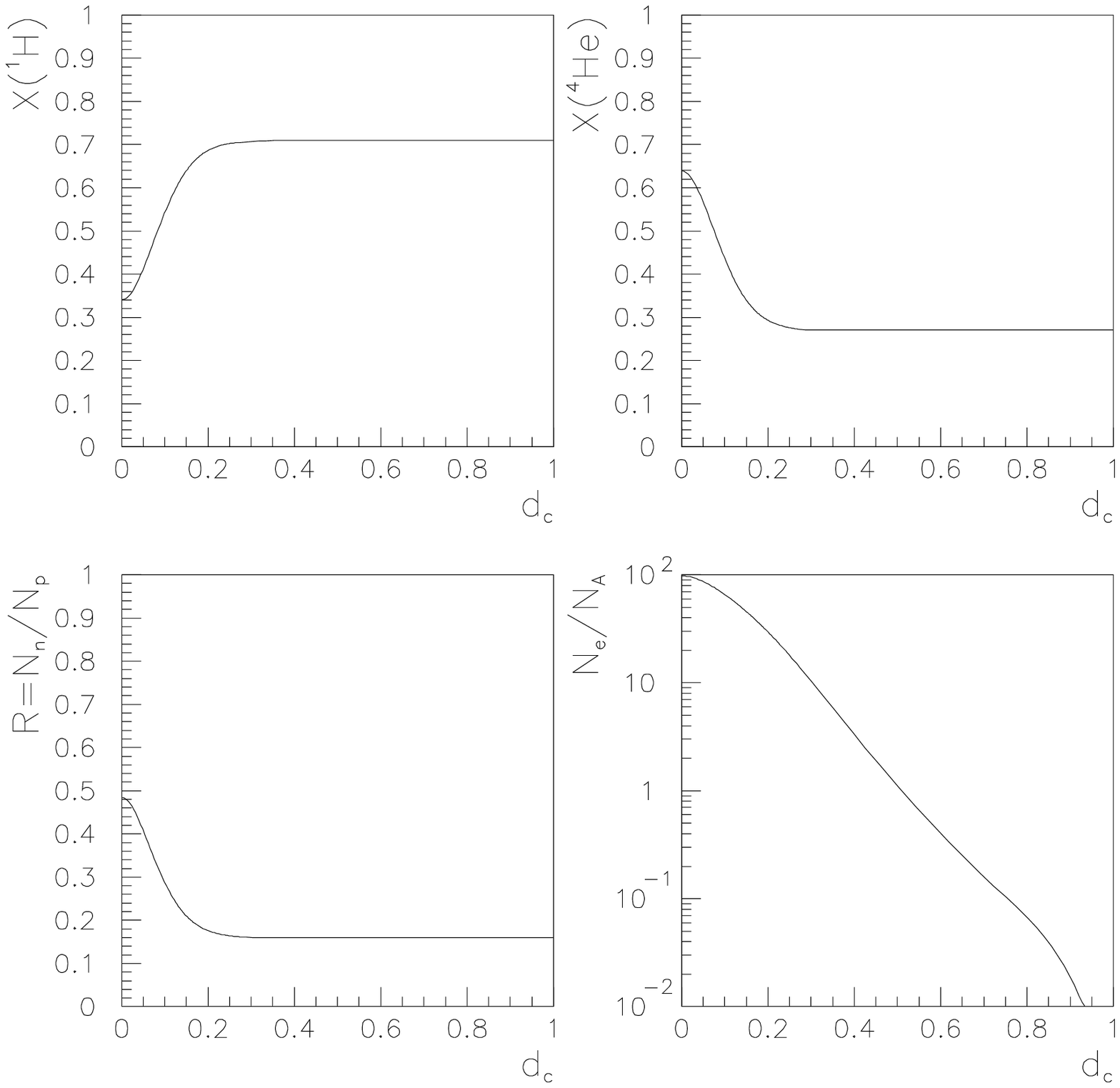,angle=0,width=16.2cm,height=16.2cm}}
\caption{The isotopic abundances $X(^1$H) and $X(^4$He), the ratio 
  $R=N_n/N_p$ and the electron density $N_e$ (in units of $N_A$) as
  functions of the distance to the solar center $d_c$ (in units of the
  solar radius). The data is taken from Ref.~[5].} 
\label{sun_data}
\end{center}
\end{figure}

\sni%
The isotopic abundances $X(^1$H) and $X(^4$He), the ratio $R$ and the
electron density $N_e$ are shown as functions of the distance to the
solar center $d_c$ (in units of the solar radius) in
Fig.~\ref{sun_data}.  In the following we discuss the subtleties that
arise due to the fact that $R$ is a function of the distance to the
solar center.

\spa%
In order to find the adiabaticity parameter $\gamma_N$ [see
Eq.~(\ref{gamma_N})] we have to calculate

\beq
\frac{d{\theta}_N}{dx} = 
     {\left(\frac{d{\tan 2\theta}_N}{d{\theta}_N}\right)}^{-1}
      \left(\frac{d{\tan 2\theta}_N}{dx}\right).
\eeq

\sni%
Taking the derivative with respect to $x$ (henceforth denoted by a
prime) we find that

\beqs
\frac{d{\tan 2\theta}_N}{dx}
&=&  \frac{d}{dx} \left(\frac{\Delta \sin 2\theta + B}
                             {\Delta \cos 2\theta - A} \right) \nonumber \\
&=&  \frac{B'(\Delta \cos 2\theta -A) - (\Delta \sin 2\theta + B)(-A')}
                  {(\Delta \cos 2\theta - A)^2}.
\eeqs

\sni%
Computing

\beqs
B' &=& 4E \sqrt{2} G_F N_e' \left[\epsilon_e 
     + 2 \epsilon_u + \epsilon_d 
     + \left(\epsilon_u + 2\epsilon_d \right) R(x) \right] + \nonumber \\
   &~& 4E \sqrt{2} G_F N_e  \left[ \left(\epsilon_u + 2\epsilon_d \right) 
            R'(x) \right]
\label{B-prime}
\eeqs

\sni%
one can see that {\it a priori} $B'$ is not proportional to $A'$ as in
the case of purely leptonic NP, due to its dependence on $R(x)$.
But to a very good approximation the second term in
(\ref{B-prime}) can be neglected, since the change of the ratio $R$
between neutron and proton density is much smaller than the change of the 
electron density, i.e. $R'(x) \ll N_e'/N_e$ (see Fig.~\ref{sun_data}). Then 
$B'$ does not depend on $N_e$, but only on $N_e'$ (like $A'$) and we obtain 
that

\beq
B=2 \epsilon(R) A \step \mbox{and} \step B'=2 \epsilon(R) A',
\eeq

\sni%
where $\epsilon(R)$ is defined in Eq.~(\ref{eps_R}). Hence we have
recovered the same formal relation between the parameters $A$ and $B$
as in the case of purely leptonic NP, only that now the proportionality
factor depends on the distance $x$ traveled by the neutrino. For the
adiabaticity parameter $\gamma_N$, which is defined at the resonance,
this implies that we can take over our previous result
(\ref{gamma_N}) (since now $B' A=B A'$) provided that we replace
$\epsilon_e$ by $\epsilon(R_{res})$:

\beq
\gamma_N(R) = \gamma |1+2\epsilon(R_{res}) \cot 2\theta|^2
\eeq

\sni%
The position of the resonance for a neutrino produced in the center of
the sun depends on its energy. We have to compute $R_{res}$ as a
function of the critical density which is given by

\beq
N_e^{crit}=\frac{\Delta \cos 2\theta}{2\sqrt{2} G_F (E_\nu)_{prod}}.
\label{Nc_E}
\eeq

\begin{figure}[\figpos]
\begin{center}
\mbox{\epsfig{figure=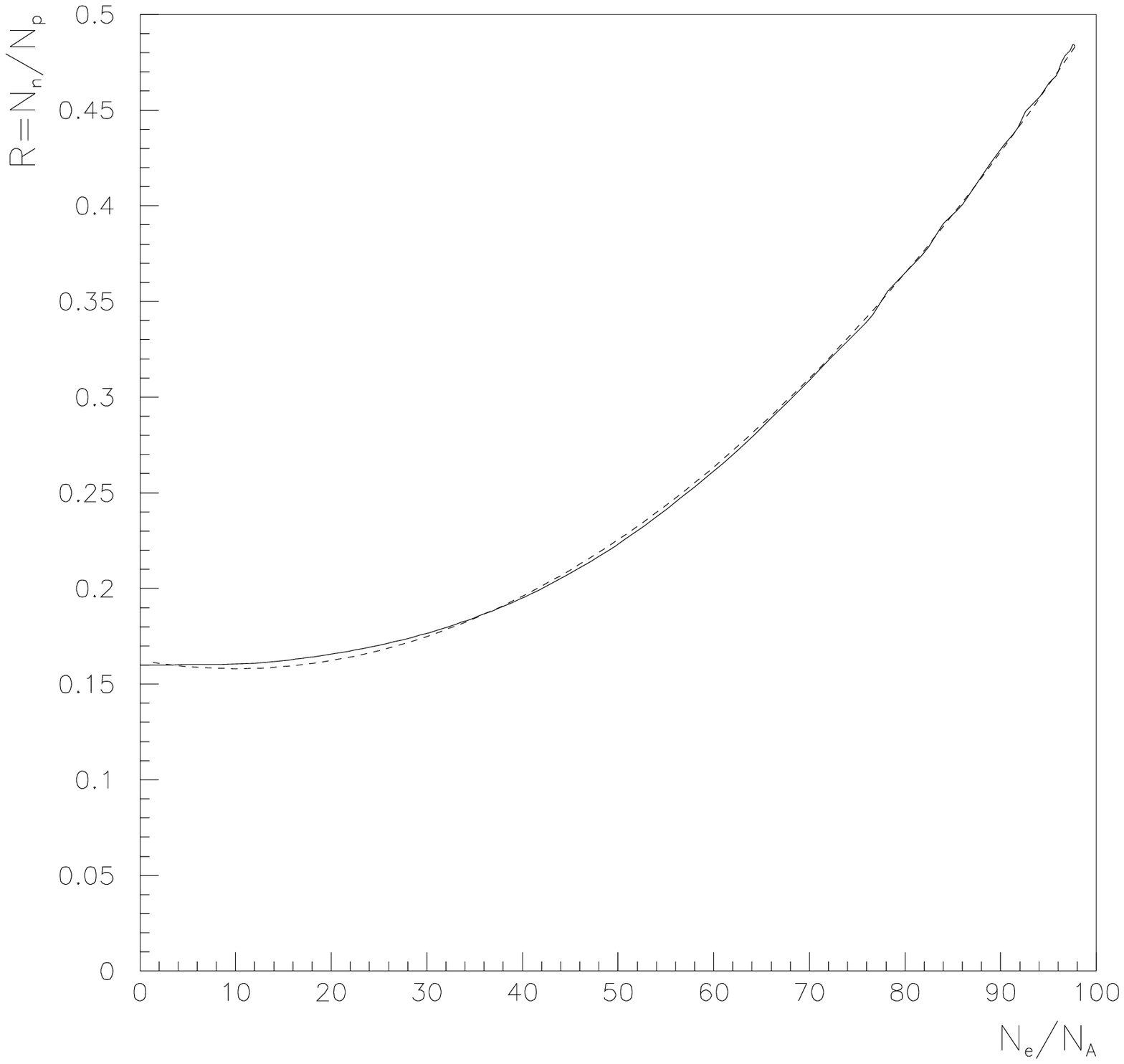,angle=0,width=16.2cm,height=16.2cm}}
\caption{The ratio $R=N_n/N_p$ as a function of the
  electron density $N_e$ (in units of $N_A$). The solid curve
  corresponds to the data from Ref.~[5] and the dashed curve
  represents the fit to a parabola.}
\label{R_Ne}
\end{center}
\end{figure}

\sni%
Thus the new adiabaticity parameter $\gamma_N(R)$ introduces an
additional energy dependence which is however not large since $R <
\frac{1}{2}$. Fig.~\ref{R_Ne} shows $R$ as a function of the electron
density $N_e$.  The dashed curve is a fit to the data points from
Ref.~\cite{Bahcall} by a parabola

\beq
R(y) \simeq 0.1624 - 0.0851 y + 0.4227 y^2,
\label{R_y}
\eeq

\sni%
where $y \equiv N_e/(100 N_A)$ and $N_A = 6.023 \times 10^{23}$ is the
Avogadro number.  

\spa%
The effects of NP neutrino interactions enter the survival probability
$P_N(E)$ [see (\ref{Pee})] via the crossing probability $P_c$ [through
$\gamma_N(R)$] and via the matter mixing $\cos 2\theta_N$ [see
(\ref{cos-theta_N})]. For $P_c$ we have to calculate $B$ (and
therefore $R$) at the resonance which gives rise to the additional
energy dependence as discussed above. The factor $\cos 2\theta_N$ in
(\ref{Pee}) has to be taken at the production point of the neutrino.
For the intermediate and high-energy neutrinos, which are mainly produced 
close to the solar center, we have $R_{prod} \simeq 0.4-0.5$. Only for 
the low-energy neutrinos, a substantial fraction
is produced with $R_{prod} < 0.4$.

\spa%
To summarize: To a good approximation all FCNC-effects due to quarks
on the neutrino propagation may be accounted for by replacing
$\epsilon_e$ with $\epsilon(R_{res})$ in the crossing probability
$P_c$ and with $\epsilon(R_{prod})$ in $\cos 2\theta_N$. The change in
$P_c$ introduces a further energy dependence which is due to the fact
that $R_{res}$ depends on the neutrino production energy.


\section{Solar Neutrino Experiment and \FCNC in the Quark Sector}
\label{quarkNPexps}

In this section we present the MSW-contours and the combined allowed
regions for the three types of solar neutrino experiments in the
presence of \FCNC in the quark sector. We have used the same
method as described previously, only that now we have included the
necessary corrections due to the energy dependence of the ratio
$R=N_n/N_p$. We have investigated the three cases of having

\begin{itemize}
\item{only $\nu-u$ FCNC $(\varepsilon \equiv \epsilon_u \ne
    \epsilon_d = 0) \To \epsilon(R)=\varepsilon (2 + R)$}
\item{only $\nu-d$ FCNC $(\varepsilon \equiv \epsilon_d \ne \epsilon_u
    = 0) \To \epsilon(R)=\varepsilon (1 + 2R)$}
\item{both $\nu-u$ FCNC and $\nu-d$ FCNC $(\varepsilon \equiv
    \epsilon_d = \epsilon_u \ne 0) \To \epsilon(R)=3\varepsilon (1 +
    R)$.}
\end{itemize}

\sni%
As an example for the MSW-contours we show in Fig.~\ref{NP_all_u},
Fig.~\ref{NP_all_d} and Fig.~\ref{NP_all_b} our results for the three
types of solar neutrino experiments with $\epsilon_u = \pm 0.05$,
$\epsilon_d = \pm 0.05$ and $\epsilon_u = \epsilon_d = \pm 0.05$,
respectively.  Comparing with the contours for
$\epsilon_e$ (see Fig.~\ref{NP_all_e}) one can see that for positive
$\varepsilon$ the changes are minor, while for negative $\varepsilon$
the region where the contours are split (at $\theta_{div}$) is
somewhat tilted.  This results from the fact that the effective
$\epsilon(R)$ of Eq.~(\ref{eps_R}) is energy dependent.  Thus the
position of the gap $\sin \theta_{div}$ is not fixed, but varies as a
function of $\Delta$. Since $N_e^{crit} \propto \Delta$, for larger
$\Delta$ also $R(N_e^{crit})$ is larger [see Eqs.~(\ref{Nc_E})
and~(\ref{R_y})]. This slightly increases $\sin \theta_{div}$
[according to (\ref{tan_div}) with $\epsilon_e$ replaced by
$\epsilon(R)$] when $\Delta$ takes larger values.  This effect is most
pronounced for $\varepsilon \equiv \epsilon_d$ (see
Fig.~\ref{NP_all_d}), since in this case the relative change
in $\epsilon(R)=\epsilon_d(1+2R)$ due to $R$ is most significant.

\spa%
As the changes in the contours due to the energy dependence of
$\epsilon(R)$ are small also the combined allowed regions for the
three cases of \FCNC in the quark sector do not differ significantly
from the results we obtained in the case of having only $\nu-\e$ FCNC.
The individual and combined allowed regions at 95\% C.L. are shown 
in Fig.~\ref{AR_NP_mu}, Fig.~\ref{AR_NP_md} and Fig.~\ref{AR_NP_mb} for
some negative $\varepsilon$. Again we have the interesting phenomena
that for a certain $\varepsilon$ there exist solutions to the SN
Problem with vanishingly small vacuum mixing. This occurs for
$\epsilon_u=\pm 0.02$, $\epsilon_d=\pm 0.03$ and
$\epsilon_u=\epsilon_d=\pm 0.01$, respectively. (For positive
$\varepsilon$ the qualitative behavior of the contours is similar to
that of $\epsilon_e > 0$ as shown in Fig.~\ref{AR_NP_pe}. For very
small $\sin^2 2\theta$ the contours coincide with those of negative
$\varepsilon$.) Note that for $\varepsilon \equiv
\epsilon_u=\epsilon_d=-0.1$ the effective $\epsilon(R) \simeq
0.35-0.45$ is so large that also the large angle solution is affected.


\section{Conclusions and Discussion}
\label{conclusions}

We have studied the effects of \FCNC on the resonant conversion of
solar neutrinos. Our main results are presented in
Fig.~\ref{AR_NP_pe}, Fig.~\ref{AR_NP_me}, Fig.~\ref{AR_NP_mu},
Fig.~\ref{AR_NP_md} and Fig.~\ref{AR_NP_mb}. We learn that the changes
to the MSW-solution could be dramatic for a NP coupling strength $G_N
\gsim 10^{-2} G_F$, in particular if the sign of $G_N$ is negative.

\spa%
There remains a question of whether couplings in the relevant range
can arise in explicit models of NP. Left-Right symmetric models
\cite{LRSM} and supersymmetric models without $R$-parity \cite{SUSY}
give rise to the purely leptonic flavor changing neutrino scattering
$\nu_e \e \to \nu_\x \e$ ($\x=\mu, \tau$).  However
model-independently these reactions are related by $SU(2)_L$-rotations
to the SM-forbidden decays $\x^\pm \to \e^\pm \e^\pm \e^\mp$.  The
experimental bounds on the branching ratios (BR) of these reactions
\cite{PDG} (BR$(\mu \to 3\e) < 10^{-12}$ and BR$(\tau \to 3\e) < 3.3
\times 10^{-6}$) imply that $\epsilon_e \lsim 10^{-6}$ for $\x=\mu$
and $\epsilon_e \lsim 10^{-3}$ for $\x=\tau$.  Since we found in
Section~\ref{lepNPexps} that for $\epsilon_e \lsim 10^{-3}$ the
modifications to the MSW-contours are very small (see
Fig.~\ref{AR_NP_pe} and Fig.~\ref{AR_NP_me}), we conclude that the
purely leptonic FCNC effects on the MSW-mechanism (in particular for
$\x=\mu$) are not likely to be significant for solar neutrinos.

\spa%
For the semi-hadronic neutrino scattering $\nu_e q \to \nu_\x q$ the
$\nu_e \to \nu_\mu$ transitions are severely constrained by the
decay $\mu \to \e \gamma$.  The experimental bound \cite{PDG}, BR$(\mu
\to \e \gamma) < 4.9 \times 10^{-11}$, implies $\epsilon_q \lsim
10^{-6}$ ruling out significant changes to the MSW-solution. However
for $\nu_e \to \nu_\tau$ oscillations the most stringent bound comes
from BR$(\tau \to \rho^0 \e) < 4.2 \times 10^{-6}$, which is much
weaker and gives $\epsilon_q \lsim 10^{-2}$.  We note that a relaxation
to our estimated bounds on $\epsilon_f$ could be achieved by $SU(2)_L$
breaking effects. However, since we consider NP at or above the
electro-weak scale, the effective four-Fermi couplings related to
$\nu_e f \to \nu_\x f$ and $\x \to f \bar{f} \e$ differ at most by a
factor of a few.

\spa%
We conclude that for $\nu_e q \to \nu_\tau q$, the strength of the
coupling in NP models could be in the range where it gives interesting
effects for solar neutrinos. Supersymmetry without $R$-parity is an
example of a model where such a coupling (for $q=d$) exists. In view
of the interesting results presented in Ref.~\cite{KB97} and in this
work, a detailed analysis of the bounds on flavor changing and new
flavor diagonal neutrino interactions in such models is called for.

\spa%
\spa%
\begin{center}
\bf Acknowledgments
\end{center}

\sni%
I am indebted to Y. Nir and Y. Grossman for many useful discussions
and valuable comments on the manuscript. I also thank E. Nardi for his
comments and J.N. Bahcall for useful communications.


\spa%
\spa%
{}


\putMSW{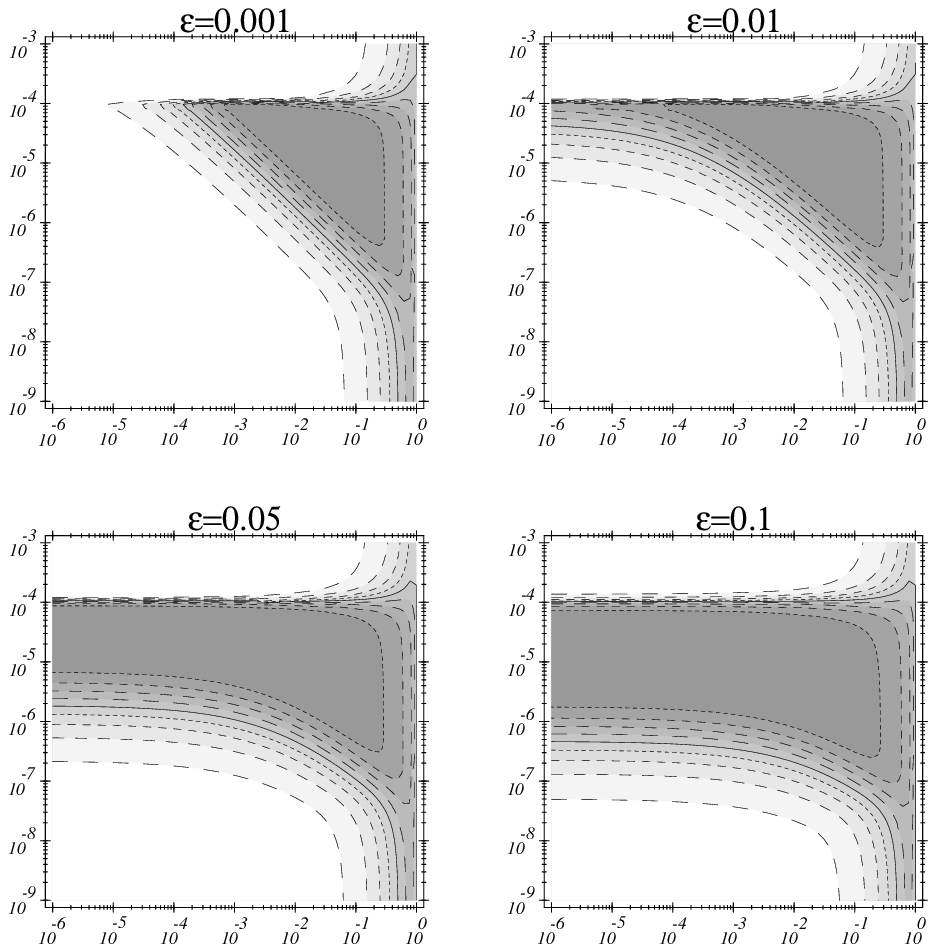}{The MSW-contours for $\varepsilon = \epsilon_e > 0$
  at one discrete energy ($E_\nu = 7$~MeV). The shading indicates the
  value of the survival probability $P_N(\nu_e \to \nu_e)$ in the
  $\sin^2 2\theta - \Delta$ plane: White corresponds to $0.9
  \le P_N \le 1.0$ and the darkest area corresponds to $0.0 \le P_N
  \le 0.1$.}{\heightAR}{\widthAR}{\spaAR}

\putMSW{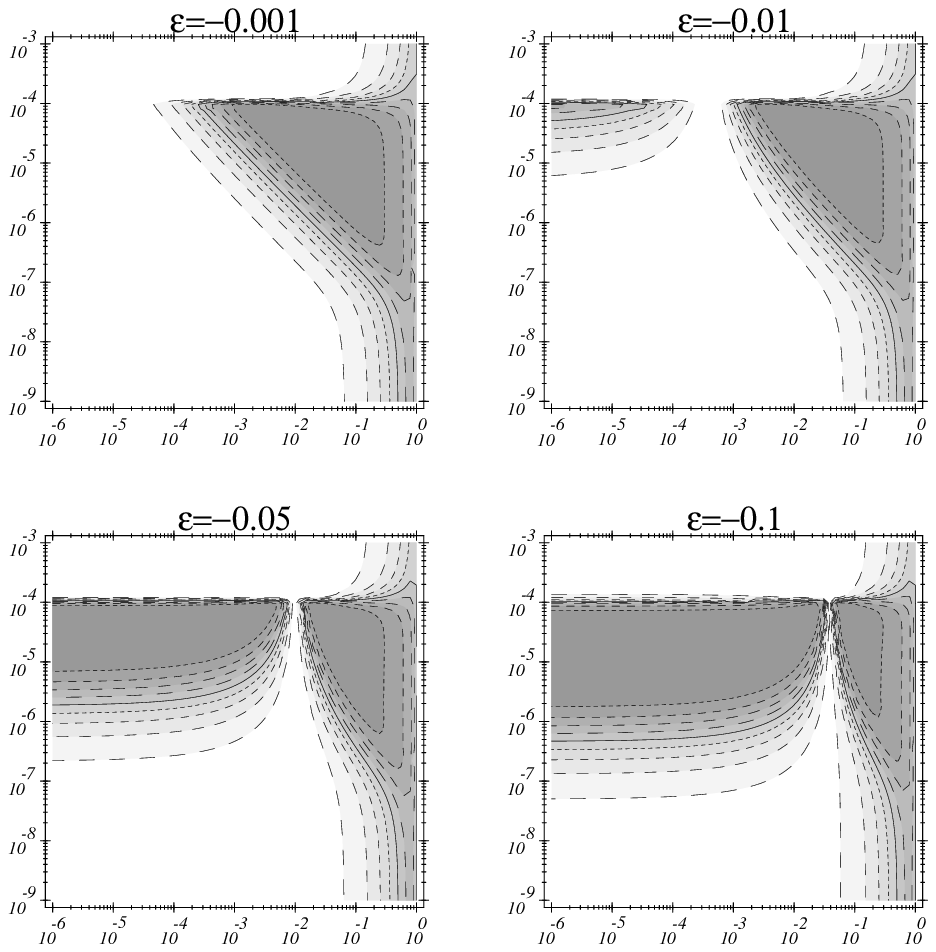}{The MSW-contours for $\varepsilon = \epsilon_e < 0$
  at one discrete energy ($E_\nu = 7$~MeV). The shading indicates the
  value of the survival probability $P_N(\nu_e \to \nu_e)$ in the
  $\sin^2 2\theta - \Delta$ plane: White corresponds to $0.9
  \le P_N \le 1.0$ and the darkest area corresponds to $0.0 \le P_N
  \le 0.1$.}{\heightAR}{\widthAR}{\spaAR}

\putMSW{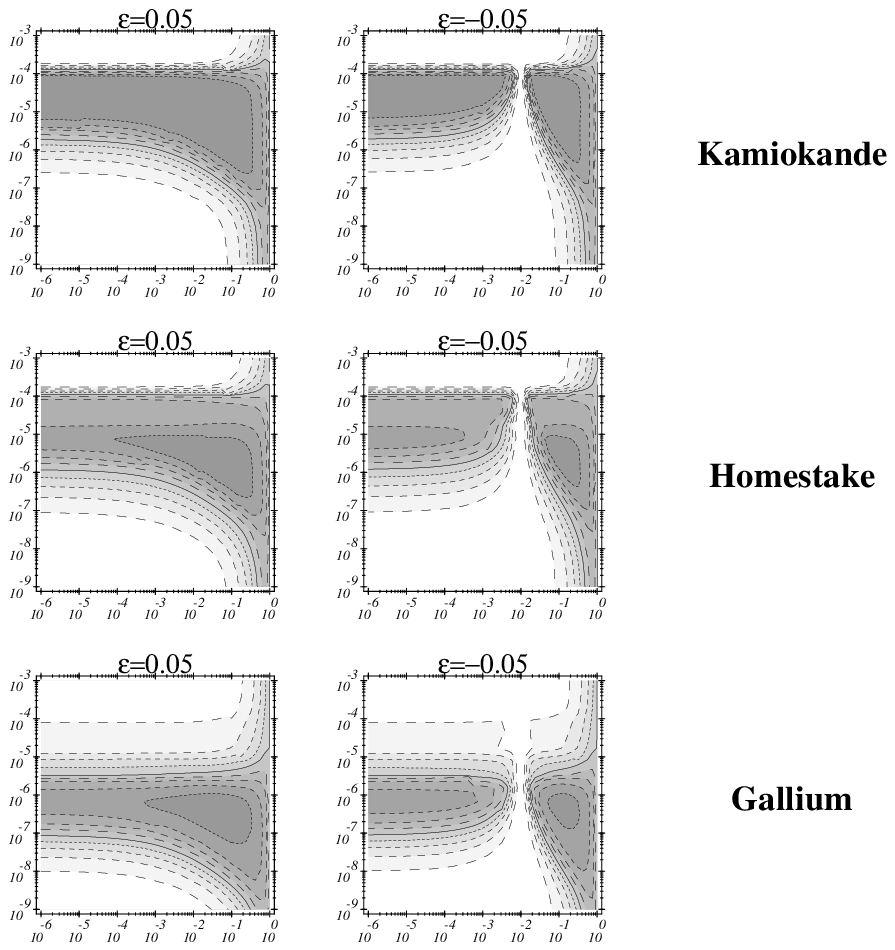}{The MSW-contours for the three types of SN
  experiments with $\varepsilon \equiv \epsilon_e = \pm 0.05$. The
  shading indicates the value of the suppression ratio $\rho$ in 
  the $\sin^2 2\theta - \Delta$ plane: White corresponds to $0.9
  \le \rho \le 1.0$ and the darkest area corresponds to $0.0 \le \rho
  \le 0.1$.}{\widthCont}{\heightCont}{\spaContour}

\putMSW{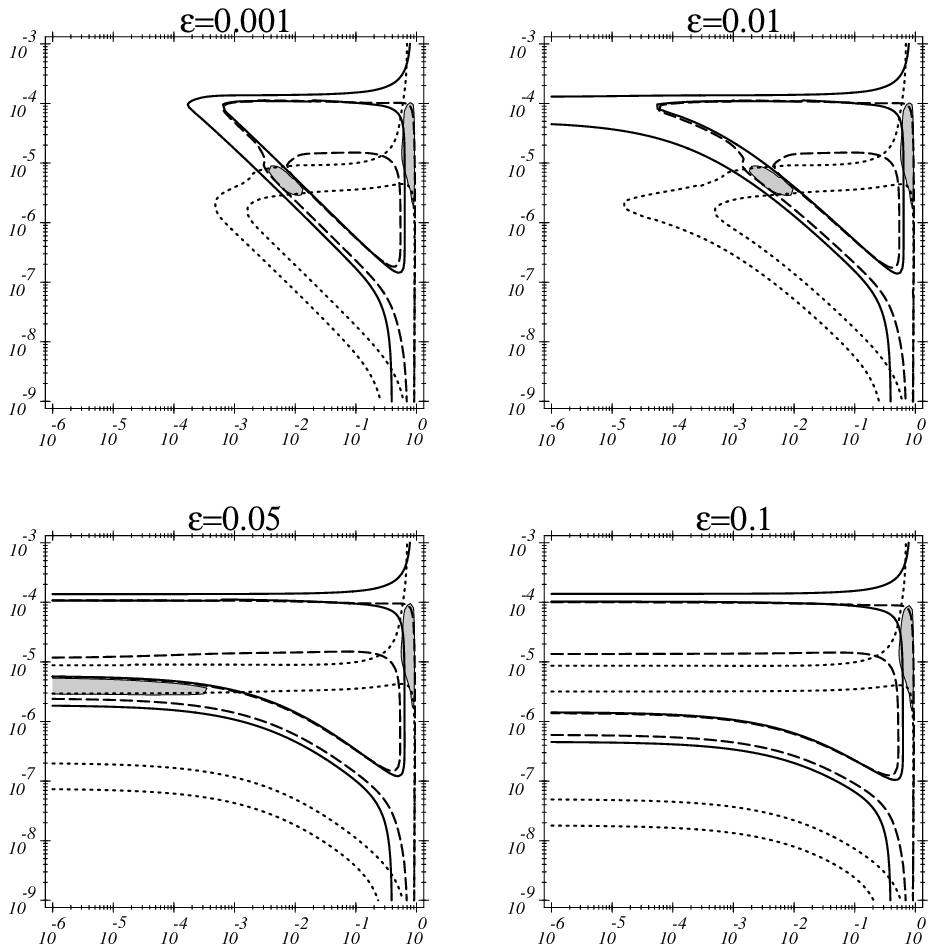}{The combined allowed regions for the solar neutrino
  experiments with $\varepsilon=\epsilon_e > 0$.  The dotted contours
  correspond to the combined gallium experiments, the dashed contours
  to the Homestake experiment and the solid contours to the Kamiokande
  experiment. The shaded areas indicate the 95\% C.L.  combined
  allowed regions in the $\sin^2 2\theta - \Delta$  
  plane.}{\widthAR}{\heightAR}{\spaAR}

\putMSW{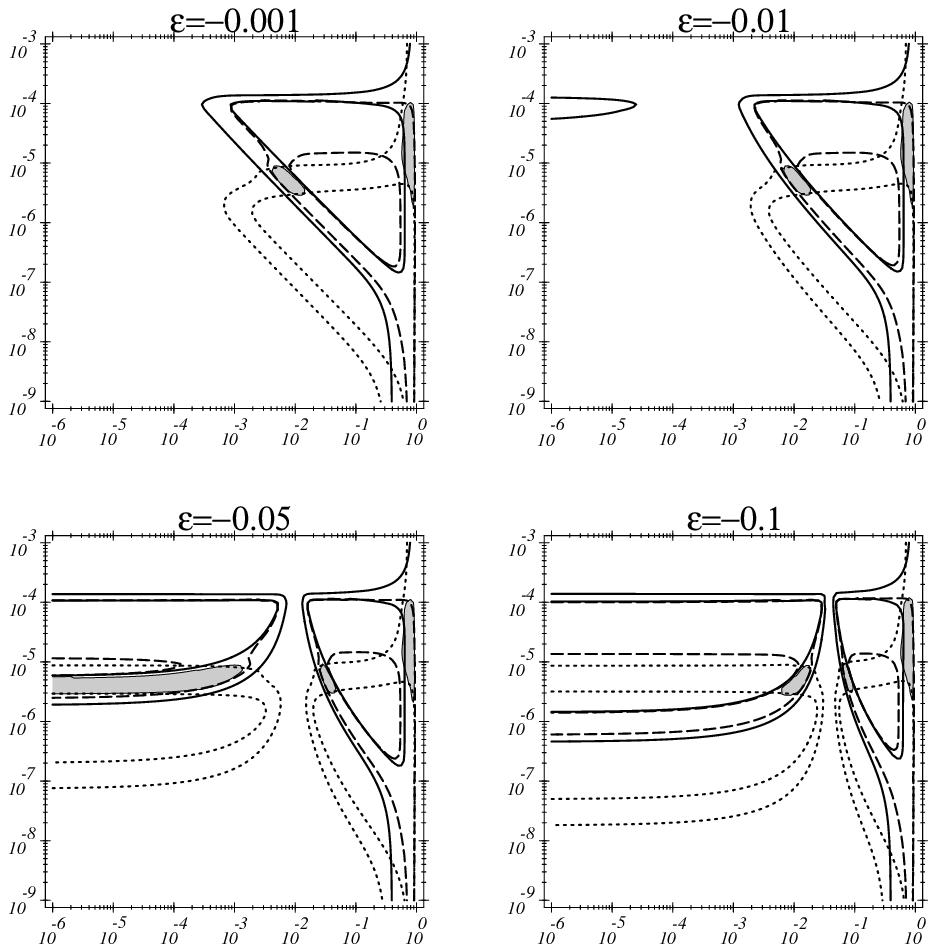}{The combined allowed regions for the solar neutrino
  experiments with $\varepsilon=\epsilon_e < 0$.  The dotted contours
  correspond to the combined gallium experiments, the dashed contours
  to the Homestake experiment and the solid contours to the Kamiokande
  experiment. The shaded areas indicate the 95\% C.L.  combined
  allowed regions in the $\sin^2 2\theta - \Delta$
  plane.}{\widthAR}{\heightAR}{\spaAR}

\putMSW{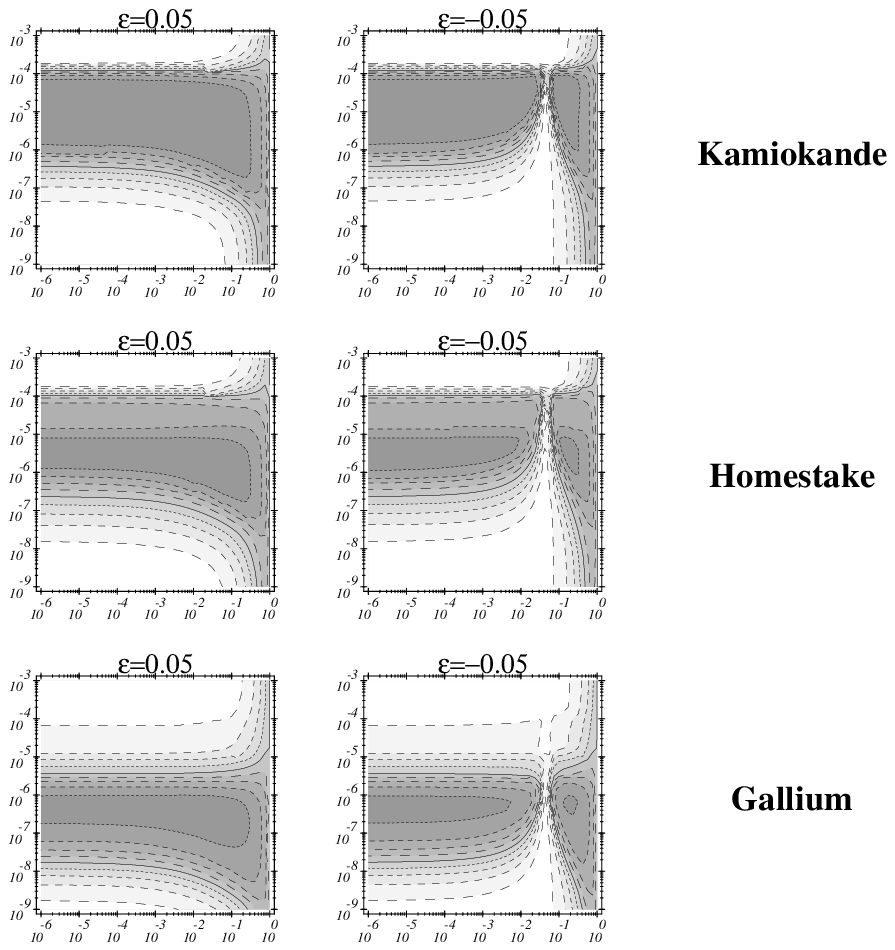}{The MSW-contours for the three types of SN
  experiments for $\varepsilon \equiv \epsilon_u = \pm 0.05$ and
  $\epsilon_d = 0$. The shading indicates the value of the suppression
  ratio $\rho$ in the $\sin^2 2\theta - \Delta$ plane: White
  corresponds to $0.9 \le \rho \le 1.0$ and the darkest area
  corresponds to $0.0 \le \rho \le
  0.1$}{\widthCont}{\heightCont}{\spaContour}

\putMSW{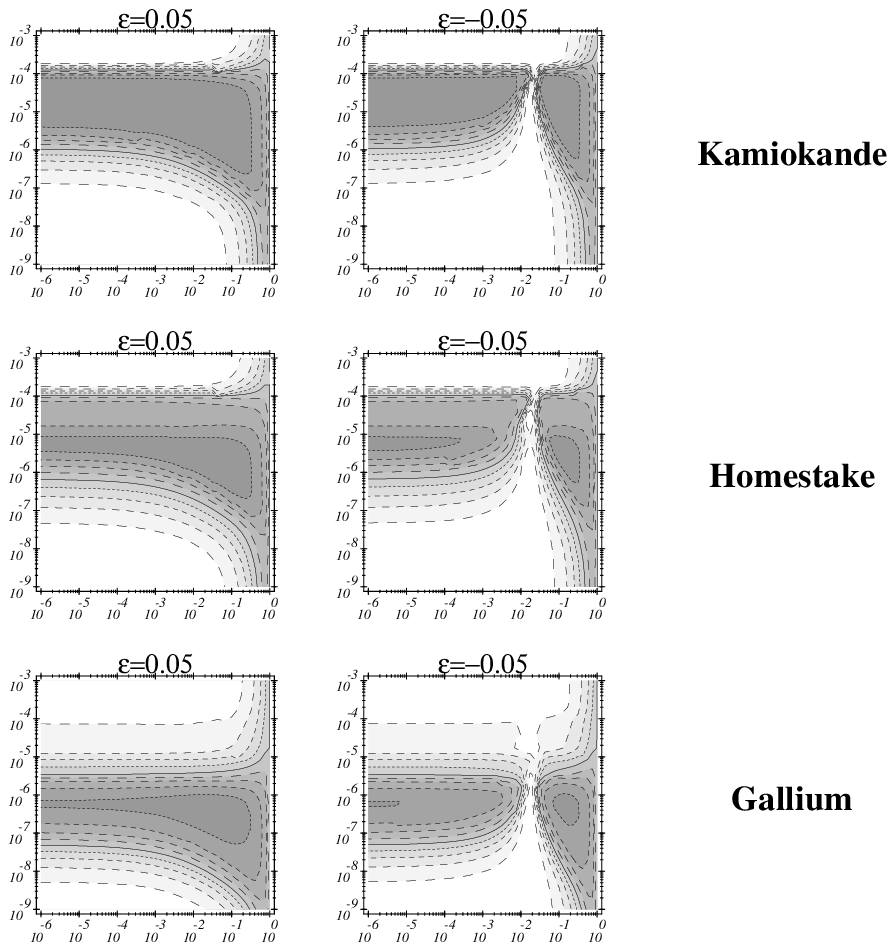}{The MSW-contours for the three types of SN
  experiments for $\varepsilon \equiv \epsilon_d = \pm 0.05$ and
  $\epsilon_u = 0$. The shading indicates the value of the suppression
  ratio $\rho$ in the $\sin^2 2\theta - \Delta$ plane: White
  corresponds to $0.9 \le \rho \le 1.0$ and the darkest area
  corresponds to $0.0 \le \rho \le
  0.1$}{\widthCont}{\heightCont}{\spaContour}

\putMSW{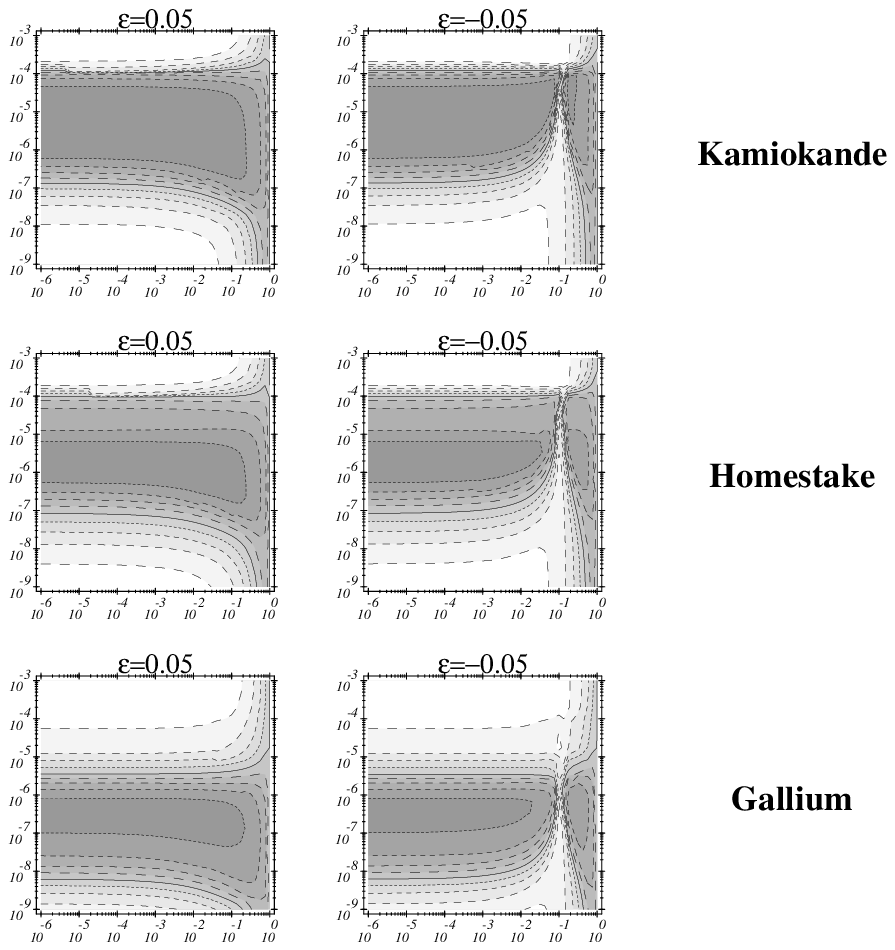}{The MSW-contours for the three types of SN
  experiments for $\varepsilon \equiv \epsilon_u = \epsilon_d = \pm
  0.05$. The shading indicates the value of the suppression ratio
  $\rho$ in the $\sin^2 2\theta - \Delta$ plane: White
  corresponds to $0.9 \le \rho \le 1.0$ and the darkest area
  corresponds to $0.0 \le \rho \le
  0.1$}{\widthCont}{\heightCont}{\spaContour}

\putMSW{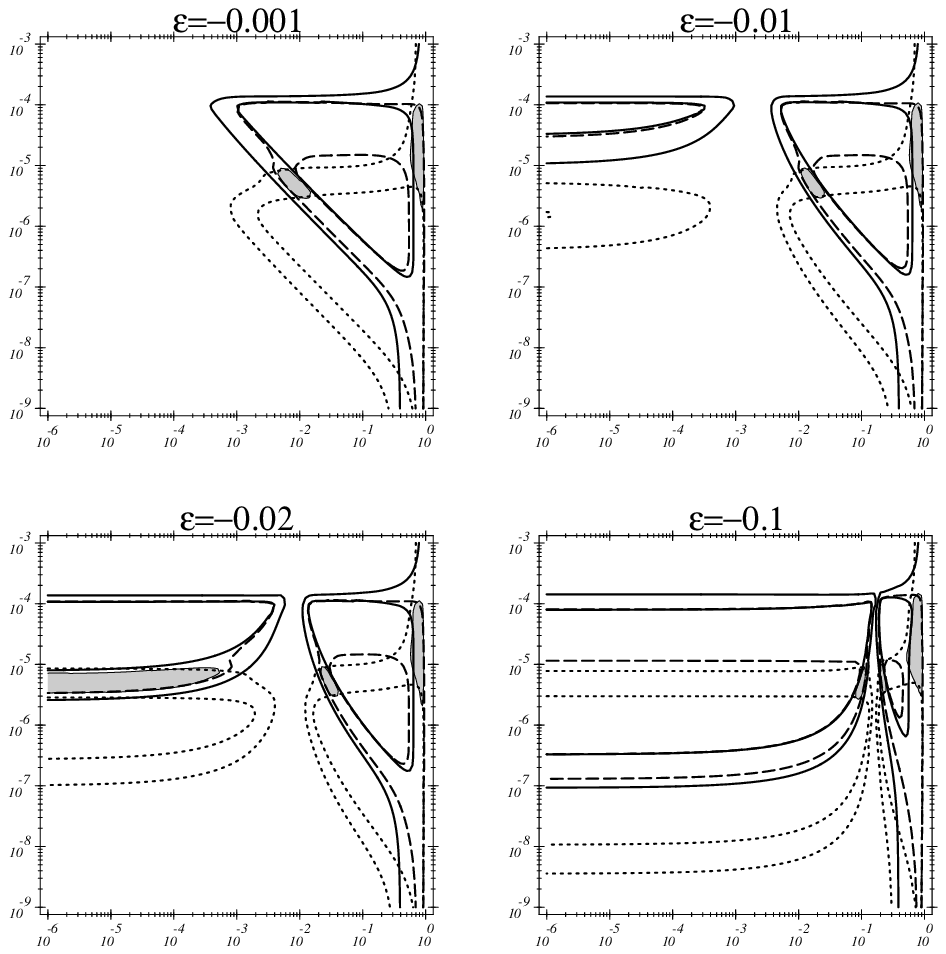}{The combined allowed regions for the solar neutrino
  experiments with $\varepsilon \equiv \epsilon_u < 0$ and $\epsilon_d
  = 0$. The dotted contours correspond to the combined gallium
  experiments, the dashed contours to the Homestake experiment and the
  solid contours to the Kamiokande experiment. The shaded areas
  indicates the 95\% C.L.  combined allowed regions in the $\sin^2
  2\theta - \Delta$ plane.}{\widthAR}{\heightAR}{\spaAR}

\putMSW{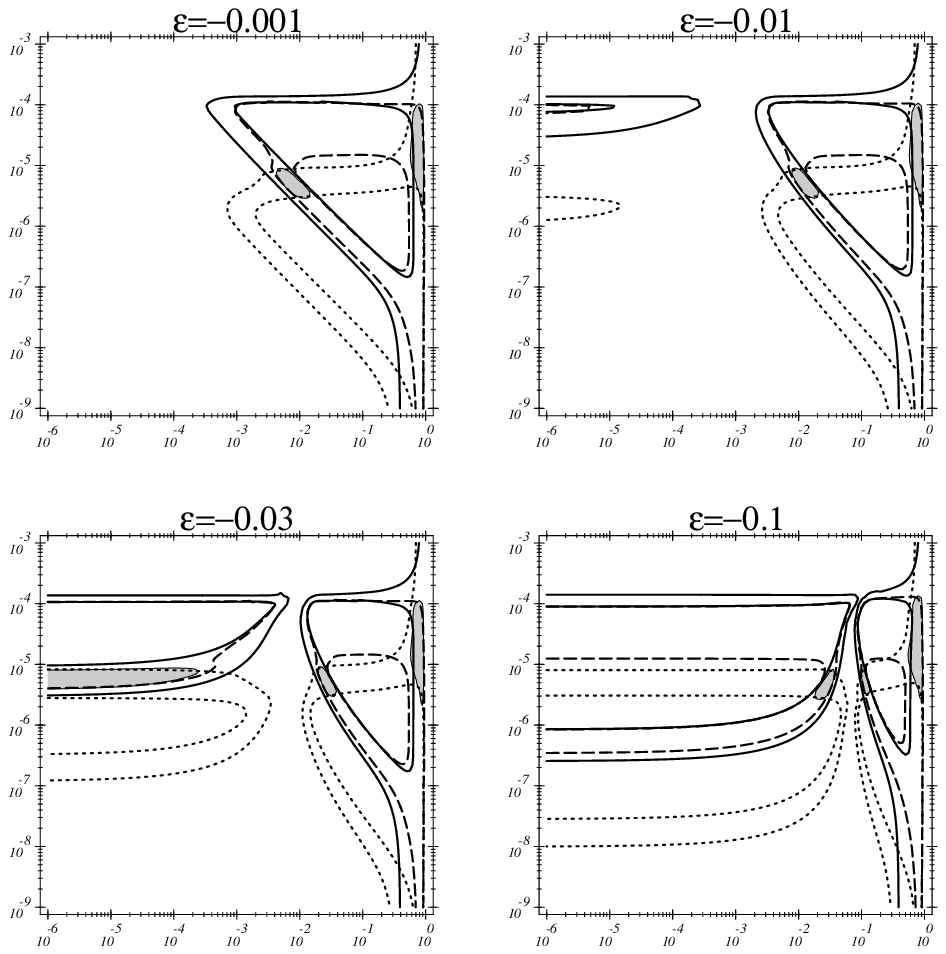}{The combined allowed regions for the solar neutrino
  experiments with $\varepsilon \equiv \epsilon_d < 0$ and $\epsilon_u
  = 0$. The dotted contours correspond to the combined gallium
  experiments, the dashed contours to the Homestake experiment and the
  solid contours to the Kamiokande experiment. The shaded areas
  indicates the 95\% C.L.  combined allowed regions in the $\sin^2
  2\theta - \Delta$ plane.}{\widthAR}{\heightAR}{\spaAR}

\putMSW{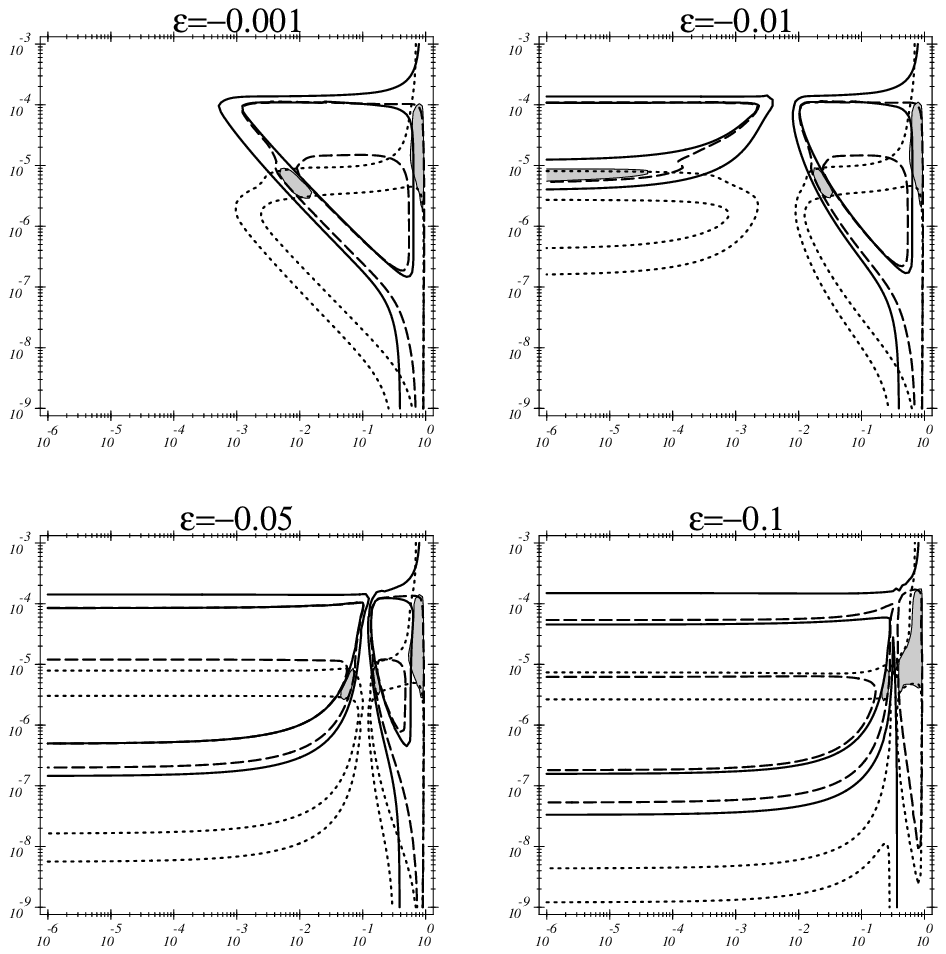}{The combined allowed regions for the solar neutrino
  experiments with $\varepsilon \equiv \epsilon_u = \epsilon_d < 0$.
  The dotted contours correspond to the combined gallium
  experiments, the dashed contours to the Homestake experiment and the
  solid contours to the Kamiokande experiment. The shaded areas
  indicates the 95\% C.L.  combined allowed regions in the $\sin^2
  2\theta - \Delta$ plane.}{\widthAR}{\heightAR}{\spaAR}

\end{document}